\documentstyle[pra,aps,amsfonts,epsfig]{revtex}

\newcommand{\mo}{\omega_-}
\newcommand{\po}{\omega_+}
\newcommand{\xo}{\omega_x}
\newcommand{\yo}{\omega_y}
\newcommand{\Om}{\Omega}
\newcommand{\vp}[1]{\varphi_{{#1}0}}

\begin{document}

\draft \title{Vortex stabilization in a small rotating  asymmetric
     Bose-Einstein condensate} \author{Marion Linn,$^{1}$
     Matthias Niemeyer,$^{1}$ and Alexander L.~Fetter$^2$}
\address{$^1$Physikalisches Institut, Universit\"at Bonn, Nu\ss allee
     12, D-53115 Bonn, Germany \\$^2$Geballe Laboratory for Advanced
     Materials and Department of Physics, Stanford University, Stanford,
     CA 94305-4045}

\date{\today}

\maketitle

\begin{abstract}
    We use a variational method to investigate the ground-state phase
    diagram of a small, asymmetric Bose-Einstein condensate with respect
    to the dimensionless interparticle interaction strength $\gamma$ and
    the applied external rotation speed $\Omega$.  For a given $\gamma$,
    the transition lines between no-vortex and vortex states are shifted
    toward higher $\Omega$ relative to those for the symmetric case.  We
    also find a re-entrant behavior, where the number of vortex cores
    can decrease for large~$\Om$.  In addition, stabilizing a vortex in
    a rotating asymmetric trap requires a minimum interaction strength.
    For a given asymmetry, the evolution of the variational parameters
    with increasing $\Omega$ shows two different types of transitions
    (sharp or continuous), depending on the strength of the interaction.
    We also investigate transitions to states with higher vorticity; the
    corresponding angular momentum increases continuously as a function
    of $\Omega$.

PACS number(s): 03.75.Fi, 67.40.Vs, 32.80.Pj\\

\end{abstract}
\section{Introduction}

The first experimental creation and detection of a vortex in a dilute
trapped Bose-Einstein condensate relied on two hyperfine components in
a nonrotating symmetric trap, with an applied electromagnetic field
coupling the two components~\cite{Matt,And}. In other recent related
experiments, several groups have used rotating nonaxisymmetric
traps~\cite{Mar,Arlt,Mad,Chevy,Abo01}, focussing, respectively, on the
irrotational flow induced by the nonaxisymmetric shape~\cite{Mar,Arlt}
and on the stabilization of vortices in such
geometries~\cite{Mad,Chevy,Mad01}. Indeed, the Paris
group~\cite{Chevy,Mad01} also
measured the angular momentum of the rotating Bose condensate.  In all
these experiments, the condensate is large and generally
well-described by the Thomas-Fermi approximation~\cite{Feder1,Feder2,AF,GO}.

In contrast, the present work focuses on a small rotating
nonaxisymmetric condensate, which requires a small dimensionless
interaction strength $\gamma$. Although current experiments have
$\gamma$ $\sim 10$-$1000$, this value, in principle, can be decreased
by reducing the particle number and/or tuning the scattering length
with the help of a Feshbach resonance~\cite{MIT,Cornish}. So far, this
limit has been studied for rotationally symmetric traps with a fixed
angular momentum $L_z$ ~\cite{Butts,Wilk,Mott,Bert,Kavou}.  However,
since experiments do not fix the angular momentum and since we focus
on nonaxisymmetric traps, we work at a fixed applied rotation $\Omega$.

We
first consider a noninteracting system in a rotating anisotropic
harmonic trap.  Although the energy eigenvalues for this problem have
been studied previously\cite{Lamb,Val,Rip,Bohr}, the relevant
structure of the corresponding quantum-mechanical eigenstates has not
been considered in detail. A subset of these low-lying eigenstates of
the anisotropic harmonic trap becomes nearly degenerate for increasing
trap rotation, and conventional perturbation theory fails (as it does
for a symmetric trap~\cite{Butts}).  Therefore, we use a variational
approach to study the ground state of the interacting system. This
method permits us to determine the phase diagram with respect to the
interaction strength $\gamma$ and the applied rotation frequency
$\Om$. Monitoring the evolution of the variational ground state
provides insight into the character of the transition to states
containing one or more vortices.  Moreover, we discuss the resulting
angular momentum carried by the condensate, which increases
quasicontinuously for rotations beyond that required for stabilization
of a few vortices.

One striking new feature of a small condensate in
a rotating anisotropic trap is the possibility of increasing the angular
momentum by elongating (which increases the moment of inertia) and
simultaneously decreasing the number of vortices (because fewer cores
can fit in the stretched form). In this way, the anisotropic
condensate mimics solid-body rotation in a qualitatively different way
from the familiar addition of more singly quantized
vortices~\cite{Butts}. Our investigations illustrate the changing
relative importance of rotational and asymmetry effects in different
areas of the $\Om$-$\gamma$ space and different values of the
anisotropy parameter
$\omega_y/\omega_x$.

The basic formalism is presented in Sec.~II along with the
variational trial wave
function.  For a fixed anisotropy, the phase diagram is discussed in
Sec.~III as  a function of the
interaction parameter and the externally applied rotation, showing
asymmetric vortex
states and in some cases re-entrant behavior.  Since angular momentum is not
conserved in an asymmetric trap, its behavior is analyzed in Sec.~IV.

\section{Basic Formalism}

In this section, we introduce the geometric  configuration, construct
the exact eigenstates of the asymmetric noninteracting Bose
condensate in a rotating anisotropic harmonic trap,  and investigate
their behavior for different
rotation speeds and asymmetries. These states provide the basis
of our variational trial function  for the
interacting system. This ansatz, along with its numerical
implementation, is  described at the end of this section.

\subsection{Hamiltonian for a rotating anisotropic harmonic trap}

To study the stationary states of a low-temperature Bose-Einstein
condensate in the rotating frame (when the trap potential becomes
time-independent), we start with the time-independent Gross-Pitaevskii
(GP) equation~\cite{Gr,Pit}
\begin{equation}
(H^{(0)} + V_H-\mu)\Psi=0.\label{GP}
\end{equation}
Here $H^{(0)}= T + V_{\rm tr} -\Omega L_z$ is the Hamiltonian for a
single particle with kinetic energy $T=-\hbar^2\nabla^2/2M$ in an
anisotropic harmonic trap $V_{\rm tr} = \frac{1}{2}M(\omega_x^2 x^2 +
\omega_y^2 y^2 + \omega_z^2 z^2)$ that rotates about the $z$ axis with
an angular velocity $\Omega$, and $V_H = g|\Psi|^2$ is the
self-consistent Hartree interaction term (the $s$-wave scattering
length $a$ determines the coupling constant $g= 4\pi a\hbar^2/M$). For
definiteness, we assume $\omega_x\le \omega_y$.

In the absence of rotation, the noninteracting Hamiltonian $H^{(0)}$
separates into a sum of three cartesian terms $H_j =
\case{1}{2}\hbar\omega_j(a_j^\dagger a_j+ a_ja_j^\dagger)$, where $j =
x, y, z$.  It is convenient to use the three oscillator lengths $d_j =
\sqrt{\hbar/M\omega_j}$ to scale the three cartesian coordinates, in
which case the harmonic-oscillator operators have the dimensionless
form

\begin{equation}
a_j = \frac{1}{\sqrt 2}\left(x_j +\frac{\partial}{\partial
x_j}\right)\quad\hbox{and}\quad a_j^\dagger  = \frac{1}{\sqrt 2}\left(x_j
-\frac{\partial}{\partial x_j}\right).\label{a}
\end{equation}
When the trap potential $V_{\rm tr}$ rotates, the term $-\Omega L_z=
-\Omega(xp_y-yp_x)$ couples $H_x$ and $H_y$, and the unperturbed
Hamiltonian becomes $H^{(0)} = H_\perp + H_z$.  Use of Eq.~(\ref{a})
shows that

\begin{eqnarray}
H_\perp &= &\frac{1}{2}\hbar\omega_x\left(a_x^\dagger a_x+
a_xa_x^\dagger\right) +
    \frac{1}{2}\hbar\omega_y\left(a_y^\dagger a_y+
a_ya_y^\dagger\right)\nonumber\\
&&+\frac{i\hbar
\Omega}{2\sqrt{\omega_x\omega_y}}\left[\left(\omega_x
+\omega_y\right)\left(a_x^\dagger a_y -a_y^\dagger a_x\right)
+ \left(\omega_x -\omega_y\right)\left(a_x^\dagger a_y^\dagger
-a_y a_x\right)\right].\label{hperp0}
\end{eqnarray}
In addition to the usual ``diagonal'' (number-conserving) terms
proportional to $a_ja_k^\dagger$, this operator also has an
``off-diagonal'' (number-violating) term proportional to $a_x^\dagger
a_y^\dagger -a_y a_x$.

As in the familiar case of the two-component Bogoliubov transformation
for a dilute Bose gas~\cite{ALF}, $H_\perp$ can be diagonalized with a
generalized Bogoliubov transformation that couples all four operators.
It is convenient to define a four-component vector $a = (a_x, a_y,
a_x^\dagger, a_y^\dagger)$; its elements obey the commutation
relations $[a_j,a_k^\dagger]= J_{jk}$, where $J$ is a diagonal matrix
with elements $(1,1,-1,-1)$. The Hamiltonian can now be written in
matrix form as $H_\perp = \frac{1}{2}\hbar\, a^\dagger {\cal H} a$,
where $\cal H$ is a $4\times4 $ hermitian
matrix given by

\begin{equation}
{\cal H}= \pmatrix{\omega_x&ic&0&id\cr-ic&\omega_y&id&0\cr 0&
-id&\omega_x&-ic\cr -id&0&ic&\omega_y},\label{ham}
\end{equation}
with
$c=(\Omega/\sqrt{\omega_x\omega_y}\,)\,\frac{1}{2}\,(\omega_x+\omega_y)$
and
$d=(\Omega/\sqrt{\omega_x\omega_y}\,)\,\frac{1}{2}\,(\omega_x-\omega_y)$.

The quadratic form $H_\perp$ can now be diagonalized with a linear
canonical transformation to a new set of four ``quasiparticle''
operators $\alpha_k$, defined by the matrix relation $a= {\cal
    U}\alpha$, where the quasiparticle operators $\alpha_k$ obey the
same boson commutation relations $[\alpha_j,\alpha_k^\dagger] =
J_{jk}$\,. The transformation matrix $\cal U$ follows from the
eigenvalue problem

\begin{equation}
{\cal H } u^{(k)} = \lambda_k J u^{(k)},\label{eigenval}
\end{equation}
where $u^{(k)}$ is the $k$th eigenvector and $\lambda_k$ is the
corresponding $k$th eigenvalue obtained from the determinantal
condition $|{\cal H} -\lambda J|=0$.  The four roots are the
eigenvalues~\cite{Lamb,Val,Rip,Bohr}

\begin{equation}
\label{fouromega}
\omega_{\pm}^2 = \omega_\perp^2 + \Omega^2 \mp
\sqrt{\case{1}{4}\left(\omega_y^2-\omega_x^2\right)^2
+4\omega_\perp^2\Omega^2},\label{freq}
\end{equation}
where $\omega_\perp^2 \equiv \frac{1}{2} (\omega_x^2+\omega_y^2)$;
these eigenvalues are identical with the classical normal-mode
frequencies\cite{Lamb}.

To understand our choice of notation $\omega_\pm$, note that the
eigenvalues reduce to $\omega_\pm^2 = (\omega_\perp\mp\Omega)^2$ for
an axisymmetric trap ($\omega_x = \omega_y=\omega_\perp$).  In the
absence of rotation ($\Omega= 0$), these two modes are degenerate.  As
is familiar from degenerate perturbation theory, applying an infinitesimal
rotation breaks the degeneracy and selects out the circularly
polarized helicity states with unnormalized wave functions $
\psi_\pm(r, \phi) \propto (x\pm iy)\,\exp(-\frac{1}{2}r^2)= re^{\pm
    i\phi} \,\exp(-\frac{1}{2}r^2)$.  In particular, the $+$ mode
rotates in the positive sense defined by the right-hand rule and its
frequency $\omega_+= \omega_\perp-\Omega$ decreases with increasing
angular velocity as is obvious when viewed in the rotating
frame\cite{Linn}.

In the general case of an anisotropic trap with $\omega_x<\omega_y$,
the two modes $\omega_\pm$ are nondegenerate even for $\Omega = 0$,
when the $+$ mode with $\omega_+ = \omega_x$ is linearly polarized
along $x$ and the $-$ mode with $\omega_-=\omega_y$ is linearly
polarized along $y$.  For nonzero rotation speed $\Omega$, the
dispersion relation (\ref{freq}) exhibits a crossover near
$|\Omega|\sim (\omega_y^2-\omega_x^2)/4\omega_\perp$ from an
anisotropy-dominated regime that favors linearly polarized states at
small $\Omega$ to a rotation-dominated regime that favors elliptically
polarized states with definite helicity at large $\Omega$.  After the
cross-over (at relatively small rotation speeds, for example $\Omega
\approx 0.05\, \omega_x$ for an asymmetry of 10\%), the
smaller positive eigenvalue $\omega_+$ decreases nearly linearly for
moderate $\Omega$; in the large-$\Omega$ regime
($\Omega/\omega_x\lesssim 1$), however, $\omega_+$ vanishes
  like $\sqrt{(\omega_y-\omega_x)(\omega_x-\Omega)}$ with an
infinite slope as
$\Omega \to \omega_x$. In contrast, the larger positive eigenvalue
$\omega_-$ increases with increasing $\Omega$.  Figure \ref{fig:eigen}
illustrates the positive eigenvalues $\omega_\pm$ as functions of
$\Omega$ for two asymmetries. For the small asymmetry
$\omega_y=1.014\, \omega_x$, the deviations of $\omega_\pm$ from the
axisymmetric case are hardly visible. For $\yo=1.1\, \xo$, the
deviation of $\xo$ from a straight line is seen at both ends of the range of
rotation speed $0\le\Omega\le\omega_x$.

The four eigenvalues in Eq.~(\ref{fouromega}) of $H_\perp$ in
Eq.~(\ref{eigenval}) can be taken to constitute a diagonal matrix
$\Lambda$ with elements $(\omega_+,\omega_-,-\omega_+,-\omega_-)$.
For the two positive eigenvalues $\omega_\pm$, the eigenvectors have
the form
\begin{equation}
u^{(\pm)}= \pmatrix{\cosh\chi_\pm\cos\theta_\pm\cr
\pm i\cosh\chi_\pm\sin\theta_\pm\cr
-\sinh\chi_\pm\sin\eta_\pm\cr
\pm  i\sinh\chi_\pm\cos\eta_\pm\cr}.
\end{equation}
These two eigenvectors satisfy the normalization condition
$u^{(j)\dagger} Ju^{(k)} = \delta_{jk}$, where $+$ and $-$ correspond
to $j$ and $ k = 1$ and $2$, respectively.  It is not difficult to
obtain explicit expressions for the (real) hyperbolic parameters
$\chi_\pm$ and for the (real) trigonometric parameters $\theta_\pm$
and $\eta_\pm$; they depend on the trap frequencies $\omega_x$ and
$\omega_y$ and on the external rotation speed $\Omega$.  Symmetry
considerations readily relate the remaining two eigenvectors (those
for the two negative eigenvalues $-\omega_\pm$) to $u^{(\pm)}$. The
resulting four eigenvectors $u^{(j)}$ with $j = 1,\cdots, 4$ form a
complete basis set and obey the normalization condition

\begin{equation}
u^{(j)\dagger} Ju^{(k)} = J_{jk}\,.\label{norm1}
\end{equation}

The transformation to the quasiparticle operators is determined by the
matrix ${\cal U}_{jk} = u_j^{(k)}$ of the four eigenvectors written in
successive columns (it is the analog of the ``modal matrix'' that
plays a central role in the theory of small oscillations of mechanical
systems about stationary configurations~\cite{FW}). In this way, the
matrix $\cal U$ satisfies the eigenvalue equation [compare
Eq.~(\ref{eigenval})]

\begin{equation}
{\cal H U} = J{\cal U} \Lambda, \label{eigen1}
\end{equation}
and the normalization (\ref{norm1}) for the eigenvectors implies that
${\cal U}^\dagger J{\cal U} = J$.  Equivalently, multiplication of the
eigenvalue equation (\ref{eigen1}) by ${\cal U}^\dagger$ gives ${\cal
    U}^\dagger {\cal H U} = J\Lambda$, showing that the transformation
matrix $\cal U$ indeed diagonalizes the hamiltonian matrix $\cal H$
with the appropriate boson metric $J$.  Correspondingly, the
quasiparticle operators follow from the matrix equation
\begin{equation}
\alpha = J{\cal U}^\dagger J a,\label{qpops}
\end{equation}
and it is easy to see that the transformed unperturbed Hamiltonian has
the expected diagonal form

\begin{equation}
H_\perp = \case{1}{2}\hbar a^\dagger {\cal H} a =
\case{1}{2}\hbar\omega_+(\alpha_+^\dagger\alpha_+
+\alpha_+\alpha_+^\dagger)+
\case{1}{2}\hbar\omega_-(\alpha_-^\dagger \alpha_-
+\alpha_-\alpha_-^\dagger)\label{hperp}
\end{equation}
that represents a set of {\em uncoupled} harmonic oscillators.
Evidently, the spectrum of allowed states has the eigenvalues
\begin{equation}\epsilon_{n_+n_-}=
\hbar(n_++\case{1}{2})\omega_+ +
\hbar(n_-+\case{1}{2})\omega_-,\label{eps}
\end{equation}
where $n_\pm$ is a nonnegative integer. As is clear from
Fig.~\ref{fig:eigen}, the set of lowest eigenvalues $\epsilon_{n_+}
=\hbar(n_++\case{1}{2})\po$ vanishes as $\Om$ approaches the
confinement limit ($\Om \to \xo$). This degeneracy precludes a simple
perturbation approach for the interacting system.

Equation~(\ref{qpops}) yields explicit expressions for the
quasiparticle operators
\begin{equation}
  \alpha_\pm = \cosh\chi_\pm\,(\cos\theta_\pm a_x \mp i\sin\theta_\pm a_y)
  +\sinh\chi_\pm\,(\sin\eta_\pm a_x^\dagger \pm i\cos\eta_\pm
  a_y^\dagger)\,,
\end{equation}
along with their adjoints. Here, the hyperbolic parameters $\chi_\pm$
and the trigonometric parameters $\theta_\pm$ and $\eta_\pm$ guarantee
the correct commutation relations for the quasiparticle operators.  In
particular, the parameters $\chi_\pm$ determine the number-violating
``Bogoliubov'' coupling between the $a$ and $a^\dagger$ operators.  As
expected from the form of Eq.~(\ref{hperp0}), these parameters vanish
for $\Omega = 0$, and $\chi_-$ remains small and positive for all
allowed $\Omega >0$, whereas $\chi_+$ is negative and decreases rapidly
as $\Omega\to \omega_x$.  For an axisymmetric trap, $\chi_\pm $ vanish
identically for all $\Omega$, and the quasiparticle
operators $\alpha_\pm$ then reduce to the familiar helicity operators
  $a_\pm = (a_x\mp ia_y)/\sqrt 2$.

  We can now analyze the noninteracting eigenstates.  The rotating
  ground state $\varphi_{00}(x,y)$ is determined from the pair of
  conditions $\alpha_\pm\varphi_{00} = 0 $; it has the form

\begin{equation}
   \varphi_{00}(x,y) = \left(\frac{ln}{\pi^2}\right)^{1/4}\,
   \exp \left[-\case{1}{2}(lx^2 +2imxy+ny^2 ) \right],
\label{gndst}
\end{equation}
where the real quantities $l$, $m$, and $n$ depend on the trap
frequencies and the rotation speed~\cite{Val}. For an axisymmetric
trap with $\omega_x=\omega_y=\omega_\perp$, the ground state has the
expected isotropic structure with $l=n = 1$ and $m = 0$.  This state
remains isotropic for all $\Omega<\omega_\perp$.

For an anisotropic trap with $\omega_x<\omega_y$, however, the
noninteracting ground state has a nontrivial phase $\propto xy$ that
represents the irrotational flow induced by the rotating
trap~\cite{JLTP,ALF,Feder2,GO,Mar,Reca01}.  For small rotation speeds and
small asymmetry, the ground-state density is indistinguishable from
that of the symmetric trap, so that it has an essentially circular
shape fitting wholly into the elliptical trap geometry. For fast
rotations, when the angular momentum per particle is considerable, the
condensate experiences a torque that stretches the ground state
density along the axis of lesser confinement (the $x$ axis in our
choice). For $10\%$ trap asymmetry, the central peak also decreases
appreciably.  As $\Omega$ approaches the smaller transverse oscillator
frequency (here, $\omega_x$), the parameter $l$ vanishes.  Thus the
oscillator confinement for the noninteracting anisotropic trap
disappears at the critical value $\Omega=\omega_x$, as is well known
from classical mechanics.

The normalized excited states are given by the familiar
harmonic-oscillator construction
\begin{equation}
   \varphi_{n_+n_-}(x,y) =
   \frac{(\alpha_+^\dagger)^{\,n_+}}{\sqrt{n_+!}}
   \,\frac{(\alpha_-^\dagger)^{\,n_-}}{\sqrt{n_-!}}\,\varphi_{00}(x,y).
\end{equation}
In the axisymmetric, static limit ($\omega_x=\omega_y,\,\Om=0$),
$\varphi_{10}$ and $\varphi_{01}$ reduce to the degenerate pair
$\psi_\pm$ discussed below Eq.~(\ref{freq}). For $\omega_x<\omega_y$,
the first excited noninteracting state $\varphi_{10}$ has an
excitation energy $\hbar\omega_+$ and represents a nonaxisymmetric
vortex with unit positive circulation and a node at the trap center.
The other noninteracting singly quantized vortex state $\varphi_{01}$
also has a node at the trap center, with higher excitation energy
$\hbar\omega_-$ and unit negative circulation.

A recent study~\cite{Butts} of a small rotating axisymmetric
Bose-Einstein condensate works at fixed angular momentum $L_z = \hbar
l$ (which is appropriate only for an axisymmetric trap).  The
resulting equilibrium configuration is then determined by minimizing
the total energy $NE_{\rm lab}(l)$ subject to the constraint of fixed
$l$.  The corresponding angular velocity $\Omega$ then follows from
the relation $\Omega = \partial E_{\rm lab}/\hbar \partial l$.

The constraint of fixed $L_z = \hbar l$ is analogous to the constraint
of fixed total $N$ in the canonical ensemble.  As in the transition to
the grand canonical ensemble, however, it is often advantageous to
eliminate the constraint by making a Legendre transformation from
fixed $l$ to fixed $\Omega$, which here merely means transforming to a
rotating frame.  Since the resulting Hamiltonian $H = H^{(0)}+ V_H$
then contains the term $-\Omega L_z$, the expectation value $\langle
L_z\rangle$ for the angular momentum as a function of $\Omega$ follows
directly from the Hellmann-Feynman theorem~\cite{Brown} $\langle
L_z\rangle = -\partial E(\Omega)/\partial \Omega$, where $E=\langle
H\rangle$ is the energy in the rotating frame.  For example, the
expectation value of the angular momentum for the noninteracting
eigenstate $\varphi_{n_+n_-}$ is simply
\begin{equation}
  L_{n_+n_-}/\hbar = - (n_++\case{1}{2})\partial \omega_+/\partial \Omega  -
  (n_-+\case{1}{2})\partial \omega_-/\partial \Omega,
\end{equation}
as follows directly from Eq.~(\ref{eps}).  Figure \ref{fig:eigen}
shows the dependence of $\omega_\pm$ on the external rotation
$\Omega$, and the resulting $L_{n_+n_-}$ also depends on $\Omega$ for
any nonzero trap anisotropy.  With the definition $L_\pm =
-\hbar\partial \omega_\pm/\partial \Omega$, we have $L_{00} =
\frac{1}{2}(L_+ + L_-)$ and $L_{10} = L_{00} + L_+$. These angular
momenta per particle for the two lowest noninteracting states
$\varphi_{00}$ and $\varphi_{10}$ are included in
Figure~\ref{fig:eigen}. For both anisotropies shown
($\omega_y/\omega_x = 1.014$ and $1.1$), $L_{00}$ remains small until
$\Omega$ approaches $\omega_x$; in contrast, $L_{10} $ rises rapidly
and linearly for small $\Omega$, remains close to one quantum of
angular momentum for most of the allowed range, and then grows rapidly
as $\Omega\to \omega_x$.

The asymmetric noninteracting excited states possess a rich structure and the
complete density distribution is needed to characterize them. As an
example, Fig.~\ref{fig:phi3dens} presents density contours of
$|\varphi_{30}|^2$ across the whole $xy$ plane, for slow, medium and
fast rotation and for two different values of $\omega_y/\omega_x>1$.
The asymmetry aligns the vortex cores along the axis of weak
confinement. At slow rotation ($\Om/\xo=0.01$), most of the condensate
density accumulates at the ends on the weak axis. For the case of
$\yo/\xo=1.1$, the depletion along the tight axis essentially splits
the condensate at the position of each vortex ``core.''
This result is plausible because this small $\Om$ is in the
asymmetry-dominated regime where $\Omega\omega_x\ll
\omega_y^2-\omega_x^2$. For medium rotation ($\Om/\omega_x=0.5$), the
effect of the asymmetry smears out, and the densities resemble
rotation-distorted symmetric trap eigenstates, with the vortex cores
located close to the trap center and surrounded by a region of nearly
constant density toward the edge. For fast rotation
($\Om/\omega_x=0.95$), the condensate stretches along the weak axis,
particularly pronounced for $\yo/\xo=1.1$. There, the non-ellipsoidal
shape of the inner most contour indicates that two of the three
overlapping vortex cores are now off-center on the $x$-axis. The
corresponding loss in angular momentum is again compensated by having  a
larger density at the ends of the weak axis. Note that the depletion
along the tight axis is here much less than for slow rotation so that
the core area is still fully surrounded by the condensate.

\subsection{Variational ansatz}

The noninteracting eigenstates now allow us to treat the interacting
problem.  With the previous dimensionless spatial variables, the GP
equation (\ref{GP}) becomes
\begin{equation}
(H^{(0)} -\mu + 4\pi \gamma \,\hbar
\sqrt{\omega_x\omega_y}\,\,\psi^* \psi)\,\psi= 0,
\label{GPdim}
\end{equation}
where $H^{(0)}=H_\perp+H_z$, with $H_\perp$ in diagonal form taken
from Eq.~(\ref{hperp}), $\gamma = Na/d_z$ is the small interaction
parameter, and $\psi$ is the condensate wave function normalized to 1.
Note that we assume a nonaxisymmetric trap rotating with a fixed
angular speed $\Omega$.  This approach describes a
``helium-bucket-like'' experiment~\cite{Mad,Chevy} and complements the
theoretical work on vortices in small axisymmetric condensates (for
example, Refs.~\cite{Butts,Kavou}), where the angular momentum (a good
quantum number) is fixed.

As is obvious from the vanishing of the noninteracting eigenfrequency
$\po (\Om)$ for $\Om \to \xo$, the lowest eigenstates $\varphi_{n0}$
become nearly degenerate for asymmetric traps, and straightforward
perturbation theory is not applicable.  Because we are interested in
small condensates with $\gamma\lesssim 1$, however, the noninteracting
eigenfunctions discussed in Sec.~II.A provide a suitable basis set for
an expansion of the planar part of the interacting condensate wave
function. Our strategy is thus to use a linear combination of the
lowest eigenstates as a trial function~\cite{Butts},
\begin{equation}
     \label{eq:varpsi}
     \psi_n (x,y) = \sum_{s=0}^n c_s \varphi_{s0}(x,y) \, ,
\end{equation}
where $n$ indicates the cutoff of the $n$th-order trial function at
the excited state $\varphi_{n0}$.  This cutoff makes the calculation
tractable; it can be chosen so that the trial function captures the
essential physics for rotation speeds at which higher excited states
do not contribute.  This trial function restricts the
region in the phase diagram that we can investigate.

We assume that the bosons are in their ground state along the axis of
rotation $\varphi_0(z) = \pi^{-1/4} \exp(-z^2/2)$. The variational
ground state follows by minimizing the free-energy functional
\begin{equation}
     \label{eq:energy}
  {E \over \hbar} =  \sum_s |c_s|^2 \omega_s^{(0)}
  + \sqrt{2 \pi} \gamma \sqrt{\xo \yo} \sum_{ijkl} c_i^* c_j^* c_k c_l
  I_{ijkl} \, ,
\end{equation}
with respect to the set of variational parameters $c_s$.  Here
$\omega_s^{(0)} = (s + {1 \over 2}) \po + {1 \over 2}(\mo + \omega_z)$
and $I_{ijkl} = \int d^2 r \varphi_{i0}^*\varphi_{j0}^*
   \varphi_{k0}^{\phantom{*}}\varphi_{l0}^{\phantom{*}}$. The variation
is constrained by the normalization condition
\begin{equation}
     \label{eq:normcond}
    \int d^2 r \,\psi^* \psi= \int d^2 r
    \sum_{ss'=0}^n  c_s^* c_{s'}^{\phantom{*}}
  \varphi_{s0}^*(x,y)
  \varphi_{s'0}^{\phantom{*}}(x,y)
     = \sum_{s=0}^n |c_s|^2 = 1 \, .
\end{equation}
In general, the parameters $c_s$ are complex, leading to
  $2 \times (n+1)$ real variational parameters. This number is reduced
by one through the norm condition~(\ref{eq:normcond}). We can eliminate
another degree of freedom by fixing the arbitrary phase of the wave
function. Therefore we have to minimize the energy~(\ref{eq:energy})
for an $n$th-order trial function with respect to $ 2 n $ independent
variables.

\subsection{Numerical implementation}

We use a well-known simplex algorithm~\cite{NumRec} to minimize the
energy~(\ref{eq:energy}) in terms of the $c_s$ as variational
variables. The algorithm basically determines the function to be
minimized with respect to $2n$ variational parameters at the corners
of a $2n+1$-dimensional simplex in parameter space.  We reduce the
number of independent variables by choosing the phase of one of the
  $c_s$ to be equal to zero and rewriting its modulus explicitly as a
function of the other variational parameters
  $|c_k|=\sqrt{1-\sum_{s \neq k}|c_s|^2}$.
The energy functional is restricted to the unit
sphere. We filled the space outside by taking the values of the energy
on the surface of the unit sphere and magnifying them with increasing
distance from the sphere. This procedure allows us to minimize in an
unrestricted space and yet ensures that the variational equilibrium
state satisfies the norm condition.  In principle, it makes no
difference which $c_k$ we exclude from the minimization procedure.
However, since we have to find the minimum with respect to many
parameters and since the energy functional has many local minima, we
swept all $k=0, \ldots ,n$ for a given point in the phase diagram.
This way, we repeated the minimization on $n+1$ different
representations of the energy and thus drastically enhanced the
likelihood of finding the true global minimum.

We monitored the reliability of our results by comparing the
contributions of higher excited states. We conclude that we have
captured all important ingredients when the distribution of weights
among the $c_s$ was not affected by including higher excited states
into the trial function. For reasons of symmetry, this test required
increasing $n$ by at least $2$.  The minimization was very robust and
reliable for most rotation speeds and interaction strengths that we
examined. The behavior of the resulting $c_s$ with increasing rotation
$\Om$ was generally very smooth apart from the transition lines
between equilibrium states with different numbers of vortices.
Another check was the convergence of the equilibrium energies for
different $n$. If a higher basis state was important, the energy
clearly decreased when it was included. If the additional dimensions
in the minimization space were irrelevant, the resulting minimal
energy was the same within the numerical errors.

\section{The phase diagram}

In this section, we determine the lowest transition lines between
variational equilibrium states in $\Omega$-$\gamma$ space for
symmetric traps and for weakly and moderately asymmetric traps. In
discussing the variational equilibrium condensate, we clarify our
criterion for the transition with increasing rotation speed $\Om$ when
the interaction and the trap geometry are fixed. Moreover, we
investigate the change in the character of the transition in different
regions of the phase diagram.

\subsection{Symmetric geometry}

First, we review the phase diagram for an axisymmetric trap that
serves as a comparison for the new features induced by the asymmetry.
The transition lines between equilibrium states are determined by
minimizing the energy~(\ref{eq:energy}) for many sets of parameters
$(\yo,\Om,\gamma)$ and comparing the resulting states. We call a state
a $q$ vortex if $\varphi_{q0}$ dominates the complete trial function
$\psi$, namely $|c_q|^2 > |c_s|^2$, for all $s \neq q$. A transition
between two states is identified by a change in the dominant weight
from $|c_q|^2$ to $|c_{q'}|^2$. This criterion works well for
``clear'' states, where one particular $|c_q|^2$ dominates strongly
over the other contributions.

 From previous theoretical studies~\cite{Butts,Gar} and from the
experiments of the Paris and MIT groups~\cite{Chevy,Abo01}, we expect that the
axisymmetric condensate accommodates the angular momentum associated
with the rotation by a sequence of transitions to states with more
vortices.  For the axisymmetric trap potential, our results agree
completely with the earlier theoretical phase diagram~\cite{Butts}.
As seen in the dot-dashed lines in Fig.~\ref{fig:PhDasym}, the first two
transition lines are strictly linear in $\gamma$ and pass through
$\Om=\omega_x$ in the noninteracting limit ($\gamma\to 0$). As
expected, the line $\Om_c(\gamma)$ for the first transition to a state
with a single vortex agrees exactly with the critical frequency
$\Om_c/\xo = 1- \gamma/ \sqrt{8 \pi}$ obtained with first-order
perturbation theory~\cite{Butts,ALF2}. Most of our data are obtained
with an $8$th-order trial function (the area very close to $\Om \to
\xo$ has been tested by using $n=10$). For the phases with up to two
vortices, only the lowest five $\varphi_{s0}$ contribute
significantly.

In order to characterize the various phases more precisely, it is
instructive to monitor the behavior of the variational parameters
along a vertical cut in the phase diagram, namely with increasing
$\Om$ for a fixed interaction strength $\gamma$.  Such a cut is
included in Fig.~\ref{fig:Occusymm} for an axisymmetric trap with
$\gamma=1$ (we measure $\Omega$ in units of $\omega_x$).  We plot the
$|c_s|^2$ starting in the no-vortex phase. Below the $\Om$ range shown
in Fig.~\ref{fig:Occusymm}, we always find $|c_0|^2=1$, confirming
that the noninteracting ground state is also the variational
equilibrium state for these slow rotations. At $\Om_c=0.80053=
1-1/\sqrt{8\pi}$, there is a sharp transition to a new state that
consists purely of the noninteracting $\varphi_{10}$ state.  This
behavior identifies the critical frequency for thermodynamic stability
of a singly quantized central vortex.  The next three transitions to
other combinations are also clearly seen, indicating a sequence of
transitions to states with an increasing number of singly quantized
vortices, each in a well-separated range of $\Om$.  Beyond the
one-vortex phase, the variational states are mixtures of various
noninteracting states.  In particular, the two-vortex phase involves
mixing with other states of two-fold symmetry
$(\varphi_{00},\varphi_{40})$ and the three-vortex phase similarly
contains other states with three-fold symmetry
$(\varphi_{00},\varphi_{60})$.  For rotation speeds higher than
$\approx 0.98\, \xo$, we are too close to the degeneracy limit $\Om
=\omega_x$ to exclude the possibility that higher-order trial
functions might change the distribution of weights. In the present
symmetric case (dot-dashed lines in Fig.~3), the order of appearance of
the different phases is the same for any vertical cut at fixed
$\gamma$ in the phase diagram.

We have included the angular momentum $L_z(\Om)=-\partial E/\partial
\Om$ in Fig.~\ref{fig:Occusymm} for $\gamma=1$.
Each transition that adds one more vortex induces a discontinuous
upward jump in the angular momentum.  Evidently, not all values of
angular momentum are allowed.  Specifically, values in the range
$0<L_z<1$ are absent, which is consistent with the character of the
equilibrium functions before and after the first transition (they
consists purely of $\vp{0}$ or $\vp{1}$, carrying exactly zero or one
quantum of angular momentum respectively). For the allowed ranges of
$L_z$, direct comparison with Ref.~\cite{Kavou} shows that we found
the same mixtures of $\varphi_{s0}$, although they fix the angular
momentum.  Since the transition lines are linear in $\gamma$, we can
also make contact with the results from Ref.~\cite{Butts}. In
particular, the first three transition frequencies indicated in their
Fig.~2 for fixed angular momentum are the same as those found here for
fixed rotation speed $\Om$.  Furthermore, the phases have the same
symmetry, the lowest two being pure $\vp{0}$ and $\vp{1}$ and thus
rotationally symmetric, whereas the equilibrium state of the $q=2$ and
$q=3$ phases have off-center vortex cores arranged to give a two-fold
or three-fold symmetric structure.

\subsection{Asymmetric geometry}

We investigate two specific trap asymmetries in detail. A weakly
distorted trap with $\yo/\xo=1.014 $ already shows some new features
and reflects a delicate balance between the symmetric
(rotation-dominated) and asymmetric (trap-dominated) influences. A
second trap geometry of $\yo/\xo=1.1$ displays more pronounced effects
of the asymmetry and thus provides a clearer picture of the basic
physics.  Figure \ref{fig:PhDasym} shows the first two transitions in
the $\gamma$-$\Om$ plane for both asymmetries, along with the
corresponding curves for the symmetric trap with $\omega_y/\omega_x =
1$.

\subsubsection{Transition lines}

First, note that the transitions for the asymmetric traps occur at
higher rotation speeds $\Omega$ than in the symmetric case; in
addition, the shift increases with increasing asymmetry. In contrast
to the linear behavior ($\Omega_c \propto \gamma$) of the symmetric
system, the transition lines curve significantly for small interaction
strengths and high rotation speeds. Most remarkably, there is a
critical threshold coupling constant $\gamma_c$ below which a ground
state with a singly quantized vortex is never favorable. This behavior
is understandable because the vortex core size decreases with
increasing interaction parameter and only fits into the trapped
condensate for not too weak interactions.
Since the semi-minor axis of the ellipsoidal trap fixes the size
available for a vortex core, a smaller asymmetry allows the
introduction of a vortex at lower $\gamma$.

Second, the term $-\Omega L_z$ in the Hamiltonian tends to favor
states with large angular momentum.  In addition to increasing the
number of vortices, a greater trap asymmetry makes the condensate more
susceptible to rotation-induced elongation, placing a greater part of
the condensate farther away from the rotation axis, increasing the
moment of inertia and hence the angular momentum. In this way, the
condensate can accommodate a higher angular momentum without
introducing vortices.

For fixed interaction strength $\gamma$ not too far above $\gamma_c$,
we even find a re-entrant region where the one-vortex phase is
followed for \emph{higher} rotation speeds by a no-vortex phase. The
tip of the one-vortex phase surrounded by the no-vortex phase for the
small asymmetry ($\omega_y/\omega_x = 1.014$) is illustrated by the
occupancies $|c_s|^2$ along a cut through that tip (at
$\gamma=0.08935$, Fig.~\ref{fig:lobe1014}). The occupancies in the
re-entrant no-vortex phase continue as if it had never been
interrupted by the one-vortex phase.  For illustration, Fig.~5 also
includes typical density contours for $\Om$ around the lobe tip. The
elongation along the horizontal $x$ axis for the vortex-free states is
pronounced. The thinner waist of the condensate on the vertical axis
results from the admixture of $\varphi_{20}$. The one-vortex state
again has only $\varphi_{10}$ as a constituent and illustrates how the
circular ring of maximal density in the symmetric case deforms to two
pronounced density peaks on the $x$ axis for this range of $\Omega$.
This re-entrant behavior reflects the singular character of the limit
$\Om \to\omega_x$ for asymmetric geometries, when the confinement parameter
$l$ in Eq.~(\ref{gndst}) tends to zero. Below the threshold interaction
strength for vortex stabilization, the density contours always
represent elongated no-vortex states; their width is smaller
  for smaller interactions strengths and increases slowly with
increased rotation.  This elongation seems to hinder vortex formation.

In order to study the detailed structure of the phases in the
asymmetric cases, we considered again the intermediate interaction strength
$\gamma =1$ and determined the occupancies $|c_s|^2$ for small 
($\omega_y/\omega_x=1.014$) and
moderate ($\omega_y/\omega_x=1.1$) asymmetry, as shown in 
Fig.~\ref{fig:Occu1014} and Fig.~\ref{fig:Occu11} respectively. As 
the applied
rotation increases, the smaller asymmetry shows phases with increasing
number of vortices, just as in the case of a symmetric trap (compare
Fig.~\ref{fig:Occusymm} for a symmetric trap). In contrast to the
re-entrant behavior for $\gamma=0.09$ (Fig.~\ref{fig:dens1014}), we
conclude that these relatively strong interactions $\gamma=1$
eliminate the effect of the (small) asymmetry, in part because the
transitions occur at slower angular velocity. For the moderate
asymmetry ($\yo=1.1\, \xo$), however, the phase diagram in
Fig.~\ref{fig:Occu11} still exhibits re-entrant behavior; thus a
$10\%$-asymmetry dominates the behavior for this interaction strength
($\gamma=1$) and precludes more than two vortex cores. Note that the
one-vortex phase continues for $\Omega/\omega_x\ge 0.99$ as if it had
not been interrupted by the two-vortex phase.

Typical density contours for the two asymmetries are included in
Fig.~\ref{fig:densasym}. Several features differ significantly from
the symmetric geometry. The no-vortex phase has a considerable
$\varphi_{20}$ admixture, producing a constriction along the $y$ axis.
In the one-vortex phase, the condensate has a central vortex, but the
elliptical trap and rotation-induced elongation deform the condensate
noticeably. For $\yo= 1.1 \, \xo$, the contributions from
$\varphi_{50}$ (and other odd states) grow as $\Omega$ approaches the
transition to two vortices, deforming the surface region because of
four vortex cores that move in from infinity. The density contours for
the interacting two-vortex state show the admixture of the
noninteracting ground state because the vortex cores are pushed
further from the center of the trap than in the symmetric case. The
small admixtures of other basis states with an even number of vortex
cores favors the accumulation of density closer to the center (outside
the core regions).

For the smaller asymmetry, the subsequent figures display three and
four separate cores, respectively. In the three-vortex phase, however,
the cores are not symmetrically distributed around the center (the
appreciable occupation of $\varphi_{20}$ enhances the two vortices
along the $x$ axis, placing the third core on the horizontal axis and
further away from the trap center).  From Fig.~\ref{fig:Occu1014},
note that the mixture of noninteracting states differs from the
symmetric case (Fig.~\ref{fig:Occusymm}), where the states contained
only noninteracting states with the same rotational symmetry.  The
last two density contours in the right column illustrate that the
re-entrant phases can indeed be classified as states with one or zero
vortices.  For such fast rotations, the condensate is very elongated
and very flat (as can be seen from the fewer density contours). It is
energetically favorable to reduce the number of vortices (which need a
wider condensate) and to compensate the loss of vorticity in the
vortex cores by expanding the condensate along the horizontal axis. In
the re-entrant $\varphi_{00}$-phase, we have several small
contributions from basis states with an even number of cores.  This
causes small ripples on a thin extended Gaussian density.  Basis
states with an odd number of cores would put a density minimum at the
center of the trap, but the dominant $\varphi_{00}$ suppresses this
tendency.

In recent experiments~\cite{Mad01}, the Paris group measured the
nucleation of vortices in their large, cigar-shaped condensate and the
corresponding angular momentum resulting from the vortices {\it
   alone}.  When ramping up the rotation, they eventually find
vortex-free states again, which is a re-entrant phenomenon similar to
what we find in our analysis for much smaller condensates.  Moreover,
they also measure the nucleation of vortices when sweeping the
asymmetry $\epsilon$ and keeping the rotation fixed, leading again to a
window of vortex stabilization.  For the small
condensate, we also determined the transition to a one-vortex state in
the $\Om$-$\yo$ plane, as illustrated for several $\gamma$ and
asymmetries up to $20\%$ (~Fig.~\ref{fig:Fasym}).  For asymmetries
larger than the rightmost end of the graphs, there is no vortex state
for rotations up to $\Om/\xo=1$.  In agreement with the $\Om$-$\gamma$
phase diagram (Fig.~\ref{fig:PhDasym}), we here see the re-entrance in
that $\Om$-$\yo$ phase diagram, leading to a qualitatively similar
restricted rotation window for stabilizing a vortex.

\subsubsection{Types of transition}

Having discussed the typical phases, we can now consider the details
of the transitions themselves.  The occupations of the separate
constituent noninteracting states (compare Fig.~\ref{fig:Occu1014} and
Fig.~\ref{fig:Occu11}) have an important new feature. There is a
smooth transition between the no-vortex and the one-vortex state for
$\gamma=1$ for both asymmetric geometries, reminiscent of a
second-order transition.  This phenomenon can already be found for a
very small asymmetry, as illustrated for $\yo= 1.001\,\omega_x$ in
Fig.~\ref{fig:Occusmall}.  The critical frequency $\Om_c = (0.80125
\pm 0.00005)\, \xo$ is a few per cent larger than $\Om_c=0.80053\,\xo$
for the symmetric trap.  Nevertheless, the density contours close to
the transition reveal that a vortex core gradually enters the
condensate along the $y$ axis (as seen in Fig.~\ref{fig:denssmall}).
This behavior is qualitatively distinct from that for the symmetric case
(for the first transition at $\gamma=1$, we examined points as close
as rotation speeds $\Delta (\Om/\xo) = \pm 10^{-7}$).  Thus we infer
that the character of the transition in asymmetric traps differs
fundamentally from the symmetric case, for we now have a cross-over
region where both $c_0$ and $c_1$ are nonzero. In this situation, the
meaning of the critical transition frequency $\Om_c$ becomes somewhat
blurred.  For both asymmetric geometries ($\omega_y/\omega_x = 1.014$
and $1.1$), this cross-over region shrinks for smaller interactions.
In fact, we find a sharp transition below $\gamma < 0.1$ for the
smaller asymmetry and below $\gamma < 0.8$ for the moderate asymmetry.
For all other subsequent equilibrium phases, we found spontaneous
jumps in the occupancies, similar to first-order transitions.

The change in character can be understood as follows: for parameters
that lead to a sharp transition, the energy functional has two
competing, well-separated main minima, one lying in the
$c_0$-dominated sector and the other in the $c_1$-dominated sector.
Indeed, depending on whether we choose $c_0$ or $c_1$ to implement the
norm condition in our minimization procedure, we find one or the other
minimum in the neighborhood of the transition. The comparison of the
energies then gives the true global minimum. The depth of these two
minima gradually changes with $\Om$; at the transition, the minimum
representing the one-vortex state becomes deeper. This picture allows
for hysteresis in stabilizing a single vortex with increasing
rotation. In fact, hysteresis is the favored explanation for the
deviation of the measured $\Om_c$ from the Thomas-Fermi predictions
(see~\cite{Chevy,Mad01,Garcia,Sinha} and references therein).
Moreover, the fact that the phase after re-entrance appear to be the
continuation of the phase {\it before} the previous transition means
that the corresponding minimum still exists and again lies below the
energy of the intervening higher-vortex state. For a continuous
transition, in contrast, the minimum energy functional must lie in a
valley connecting the two sectors. Here, the different implementations
of the norm condition lead to the same minimal state. With increasing
$\Om$, this global minimum gradually moves along this valley from the
$c_0$-dominated sector through a cross-over region to the
$c_1$-dominated sector.

 From Fig.~\ref{fig:Occu1014} for the smaller asymmetry with $\yo=1.014
\, \xo$, we can also observe that the three-vortex phase undergoes
much stronger changes across its range than other phases.  Although
there is a sharp transition to a phase with $\varphi_{30}$, an
appreciable amount of the two-vortex basis-state remains and
$\varphi_{20}$ dies out only gradually. This behavior yields a very
complicated picture for smaller interaction strengths (roughly at
$\gamma\approx 0.6$), where the $\varphi_{30}$-dominated phase disappears
completely leading to a direct transition from two to four vortices.
For even smaller $\gamma$, the two-vortex phase is followed by a
one-vortex phase before the four-vortex phase develops. In part, this
complicated picture arises from the suppression of the $\varphi_{30}$
contribution to the one-vortex phase (which in the moderately
asymmetric trap with $\yo= 1.1 \, \xo$ leads to an contribution of
$\varphi_{50}$ rather than $\varphi_{30}$ to the one-vortex phase).

\section{The angular momentum}

The asymmetry in the trap geometry breaks the cylindrical symmetry, so
that the angular momentum $L_z$ around the axis of rotation is no
longer a good quantum number. We therefore investigate the effect of
the asymmetry on $L_z$ in some detail.

The angular momentum (in units of $\hbar$) as a function of the trap
rotation $\Om$ is shown in Fig.~\ref{fig:Lz} for all three geometries
($\omega_y/\omega_x = $ 1.0, 1.014, and 1.1) for the fixed interaction
$\gamma=1$.  The intimate relation to the occupancies within the
various phases for each geometry can be seen in the corresponding
plots (Figs.~\ref{fig:Occusymm},~\ref{fig:Occu1014},
and~\ref{fig:Occu11}). The sharp kinks in $L_z$ for the symmetric trap
arise from the sudden changes in the occupancies and the different
angular momentum carried by the noninteracting states. The finite
slope of the plateaus beyond the second transition reflects the mixing
of the various noninteracting states in the variational ground state
and the  off-center positions of the vortex
cores~\cite{Butts}.

For the asymmetric rotating trap, even the noninteracting eigenstates
$\varphi_{s0}$ with a central $s$-fold vortex carry an angular
momentum different from $s$ (in fact, the angular momentum diverges
for $\Om \to \omega_x$). The divergence comes from the contribution of
the circulating quanta with positive helicity, $\, L_+ (\Om)$,
occurring in the angular momentum of every basis state $\varphi_{s0}$.
This is easily seen by expanding $L_+ = -\partial \omega_+/\partial
\Om$ for small $ \omega_x-\Omega$ and small asymmetries $\yo^2 = 1 +
2\epsilon^2$, leading to
\begin{equation}L_+ \approx  \frac{
\epsilon}{\sqrt{\omega_x-\Om}}\quad\hbox{for
$\Omega\to\omega_x$} \,.
\end{equation} In asymmetric traps, this divergence dominates the
angular momentum in all noninteracting eigenstates $\varphi_{s0}$ for
high rotation speeds and hence in any linear combinations of them.
Thus, even discrete changes in the structure of the equilibrium state
(as at a transition) will have a far less dramatic effect on the
angular momentum at large rotation speeds near $\omega_x$. Moreover,
asymmetric condensates are significantly elongated along the axis of
weaker confinement as $\Omega\to\omega_x$. The resulting
redistribution of density (and mass) with respect to the axis of
rotation produces an additional, nonquantized contribution to the
angular momentum.

Compared to the symmetric condensate, the angular momentum for the
trap with smaller asymmetry at $\gamma=1$ has its jumps smoothed over
a small but finite range of $\Omega$ because of the continuous
transition where one off-center vortex gradually moves towards the
center (see Fig.~\ref{fig:Occu1014}); in addition, the transition is
shifted toward
higher rotation speeds (see Fig.~\ref{fig:Lz}). The subsequent
sequence of phases
corresponds to an increasing number of vortex cores that cause nearly
vertical jumps just as in the symmetric case. The divergence of the angular
momentum dominates only at $\Om/\omega_x\approx 1$.

For the moderate asymmetry, the stronger admixture of $\varphi_{20}$
in the no-vortex phase yields a faster rise of the angular momentum
below the first transition. The first kink is even more tilted because
of the extended range for the transition to a one-vortex state.
Inside this one-vortex phase, there is a broad region with a pure
$\varphi_{10}$
state (up to $\Om/\omega_x \approx 0.92$). The angular momentum,
however, exceeds one because of both the initial growth of $L_+$ with
$\Om$ and the deformation of the condensate density.  The rapid rise
of $L_z$ below the second transition reflects the growing admixture of
$\varphi_{50}$.  That second transition, seen as a discontinuous
derivative in $L_z$, occurs at an angular momentum $L_z \approx 2.4$,
already exceeding the value for a symmetric trap with two vortices.
Beyond that point, the angular momentum increases smoothly with $\Om$,
even beyond the re-entrant transitions at $\Om \approx
0.9905\,\omega_x$ and $\Om \approx 0.9945\,\omega_x$ respectively (see
inset in Fig.~\ref{fig:inset}). In this limit, the divergent $L_+$ completely
dominates the behavior.

\section{Conclusion}

The behavior of a small Bose condensate in a rotating anisotropic trap
differs significantly from that of a symmetric condensate, even for
very small asymmetries.  In large part, this difference arises from the
stretching of the condensate along the axis of weak confinement, especially
for
$\Omega\to\omega_x$.
In addition, it reflects the irrotational flow induced by the
rotating confining
potential that pushes on the gas when viewed from the laboratory
frame~\cite{JLTP,Feder1,Feder2,GO}.

For any specific asymmetry and sufficiently small interaction
strength, a one-vortex state can never be stabilized, as illustrated
in Fig.~\ref{fig:PhDasym}.  For 
  somewhat larger
values of the coupling constant $\gamma$, the first transition, from a
no-vortex state to a one-vortex state, is sharp, but the subsequent
transitions are re-entrant, back to a no-vortex state with significant
elongation that carries the relevant angular momentum (see
Figs.~\ref{fig:lobe1014} and \ref{fig:densasym}).  For these weak
interactions, the transitions all occur close to
$\Omega\approx \omega_x$, so that the stretching dominates.

For larger interaction strength ($\gamma\sim 1$), the transitions
occur at lower values of $\Omega$, and they become quasicontinuous in
that the occupation $|c_0|^2$ of the noninteracting ground state
vanishes smoothly (see Figs.~\ref{fig:Occu1014}, \ref{fig:Occu11},
   and \ref{fig:denssmall}).  As in the case of a
classical fluid in a rotating elliptic cylinder~\cite{JLTP}, the
angular momentum is finite even below the transition to a state with
one vortex because of the irrotational flow induced by the rotating asymmetric
trap (see Fig.~\ref{fig:Lz}).

\acknowledgements This work was supported in part by the National
Science Foundation Grant No.~DMR-9971518 and by the Deutsche
Forschungsgemeinschaft Grant No.~SSP~1073. ALF is grateful to G.~Bertsch for
guidance on the nuclear-physics literature and to the Aspen Center for
Physics where this work was initiated.

\begin{figure}
     \begin{center}
       \caption{The positive eigenvalues $\omega_\pm$ and the
angular momenta $L_{00}$ and $L_{10}$ of the two lowest
noninteracting eigenstates $\varphi_{00}$ and $\varphi_{10}$ as a
function of the rotation speed $\Omega$ for asymmetries
$\omega_y=1.014$ (dashed lines) and
$\omega_y=1.1$ (solid lines). All quantities are in dimensionless
units (scaled with respect to
the lower oscillator frequency $\omega_x$).  }
       \label{fig:eigen}
     \end{center}
\end{figure}


\begin{figure}[ht]
\begin{center}
\end{center}
\caption{Density contours of the noninteracting $|\varphi_{30}|^2$ across
the $xy$ plane for the
traps with $\yo/\xo=1.014$ (left) and $\yo/\xo=1.1$ (right)
for $\Om/\xo =0.01, \, 0.5$
and $0.95$ (top to bottom). Distances are scaled in units of
$d_x$ in \emph{both}
directions and the width shown is $6 d_x$ across. }
\label{fig:phi3dens}

\end{figure}

\begin{figure}[htbp]
     \begin{center}
       \caption{The two lowest transition lines for $\yo= 1.0 \, \xo$
(dot-dashed lines),
$\yo= 1.014 \, \xo $ (dotted lines) and $\yo= 1.1 \, \xo$ (solid lines),
determined with trial functions up to $10$th order. The lowest transition lines
(a,b,c) represent the critical rotation $\Om_c$ for the stabilization
of a single
vortex; at the second transition lines (d,e,f) the condensate starts
to be dominated
by the $\varphi_{20}$. There are unresolved higher vortex-phases
beyond the second
transition for sufficiently large  interaction strength.}
       \label{fig:PhDasym}
     \end{center}
\end{figure}

\begin{figure}
\caption{\protect{$\omega_y=1.0\,\omega_x$, $\gamma=1:$} Occupancies
   (left scale) and $L_z$ (right scale) for axisymmetric condensate as
   function of $\Omega$ for fixed interaction $\gamma=1$. The numbers
   above the lines denote the corresponding $|c_s|^2$ and the starred
   solid line represents the angular momentum $L_z$.}
\label{fig:Occusymm}
\end{figure}

\begin{figure}[htp]
    \begin{center}
   \end{center}
\vspace{3mm}
\caption{Occupancies $|c_s|^2$ for the basis-states for $\yo=1.014\, \xo,
    \, \gamma=0.08935$, calculated with a $8$th-order trial function and
    typical density contours (at $\Om=0.99515,\,0.9952, \, 0.9953$).
    Contributions for $s>2$ are negligible. The tip of the pure
    $\varphi_{10}$-lobe cuts sharply into the no-vortex phase, which
    has an appreciable $\varphi_{20}$ admixture. Note the narrow range
    in $\Om$. The density contours are shown over 6 oscillator lengths
    $d_x$ in each direction.}
\label{fig:lobe1014}  \label{fig:dens1014}
\end{figure}

\begin{figure}
\caption{$\omega_y=1.014\,\omega_x$, $\gamma=1:$ Occupancies (left scale)
and
   $L_z$ (right scale, note upward shift). The numbers above the lines are
short for the
   correspondent $|c_s|^2$ and small contributions from $|c_6|^2 \ldots
   |c_{10}|^2$ are not separately labeled. The starred solid line
   represents the angular momentum $L_z$.  }
\label{fig:Occu1014}
\end{figure}

\begin{figure}
\caption{$\omega_y=1.1$, $\gamma=1:$ Occupancies (left scale) and $L_z$
(right scale). The numbers above the lines are short for the
correspondent
   $|c_s|^2$ and small contributions from $|c_3|^2, \, |c_6|^2 \ldots
   |c_{10}|^2$ are not separately labeled. The starred solid line
   represents the angular momentum $L_z$.}
\label{fig:Occu11}
\end{figure}


\begin{figure}[t]
     \begin{center}
  \end{center}
\caption[Density contours for various phases and small and moderate asymmetry]
{Density contours for variational equilibrium states for small (1.014
  left) and moderate (1.1 right) asymmetry and $\gamma=1$. The
  pictures represent typical states within the different phases (cf.
  Fig.~\ref{fig:Occu1014} and ~\ref{fig:Occu11} respectively).  For
  the small asymmetry, the number of vortices increases from zero to
  four ($\Om=\, 0.805,\, 0.935 , \, 0.95, \, 0.97, \, 0.982$) whereas
  in the moderate asymmetry, we find re-entrant behavior into one and
  zero-vortex states ($\Om=\, 0.87,\, 0.9, \, 0.97, \, 0.993, \,
  0.997$). The maximum width shown is 6 oscillator lengths $d_x$.}
       \label{fig:densasym}
\end{figure}

\begin{figure}[ht]
    \begin{center}
      \caption{Lines of critical rotation for the stabilization of one vortex
        versus the asymmetry, for $\gamma=0.8,1,1.5$ (left to right),
        from 8th-order trial functions. The one-vortex states are to
        the left of each graph, the vortex-free ones to the right. All
        quantities are given in dimensionless units.}
      \label{fig:Fasym}
   \end{center}
\end{figure}


\setlength{\unitlength}{1mm}
\begin{figure}[htp]
\begin{center}
\end{center}
\vspace{3mm}
\caption[First transition for a very small asymmetry]{Occupancies and
density contours for $\omega_y=1.001$,
$\gamma=1$, from an $8$th-order trial function. The occupancies for
$\varphi_{30}$ and higher are not labeled for clarity. The density
contours are taken at rotations $\Om = 0.8007,\, 0.8011, \, 0.8012,
\, 0.8013$ and are about 4 oscillator lengths $d_x$ across. }
\label{fig:Occusmall}\label{fig:denssmall}

\end{figure}

\begin{figure}[thp]
\begin{center}
\end{center}
\vspace{3mm}
       \caption[The angular momentum for $\yo/\xo = 1.0,
       \, 1.014, \, 1.1$ and $\gamma=1$] {The angular momentum
         $L_z$ in units of $\hbar$ as a function of the angular velocity
         $\Omega$ for the three geometries $\yo/\xo = 1.0, \, 1.014, \,
         1.1$ (dot-dashed, dotted and solid lines respectively). The
         inset shows the magnified re-entrance range for $\yo/\xo =1.1$.}
       \label{fig:Lz}\label{fig:inset}
\end{figure}

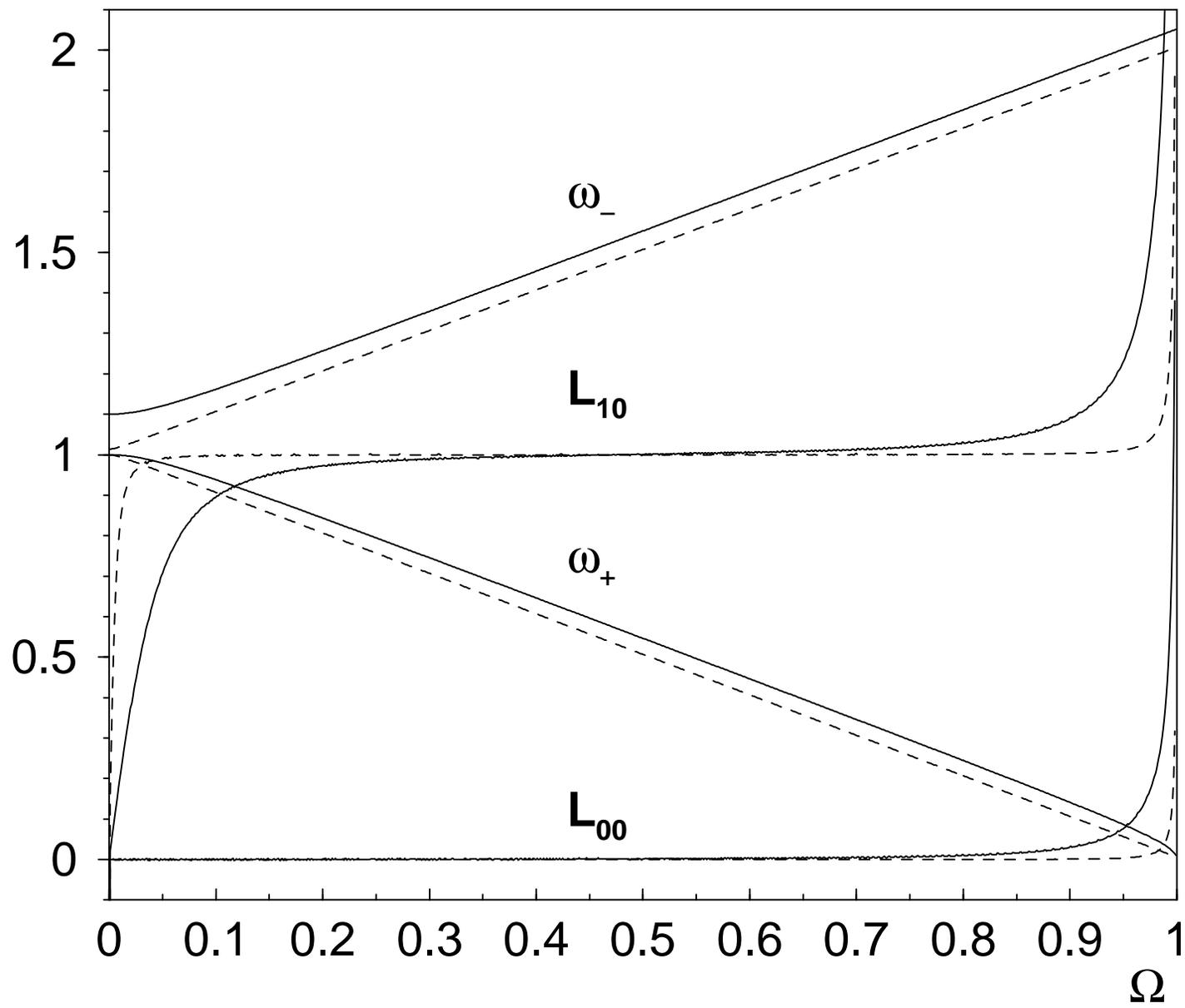
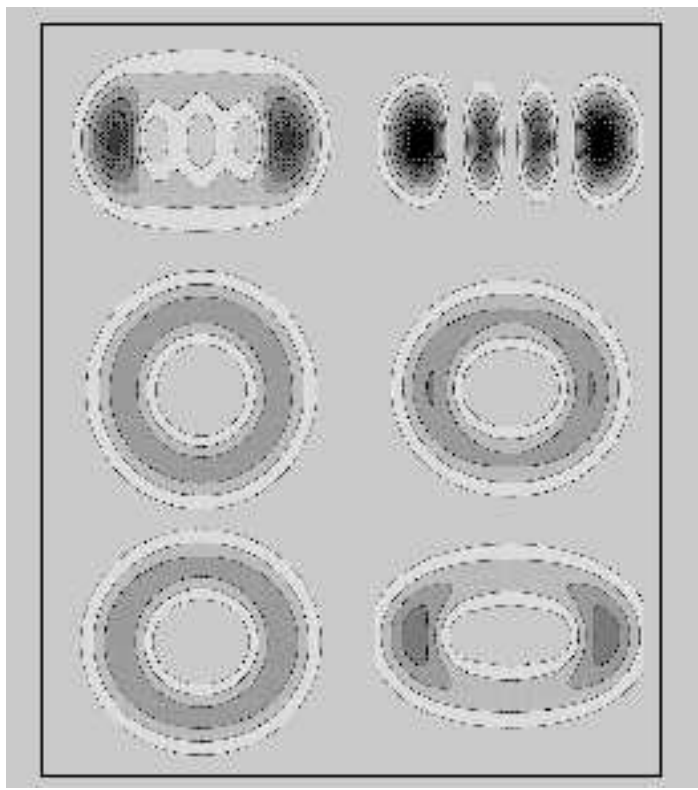
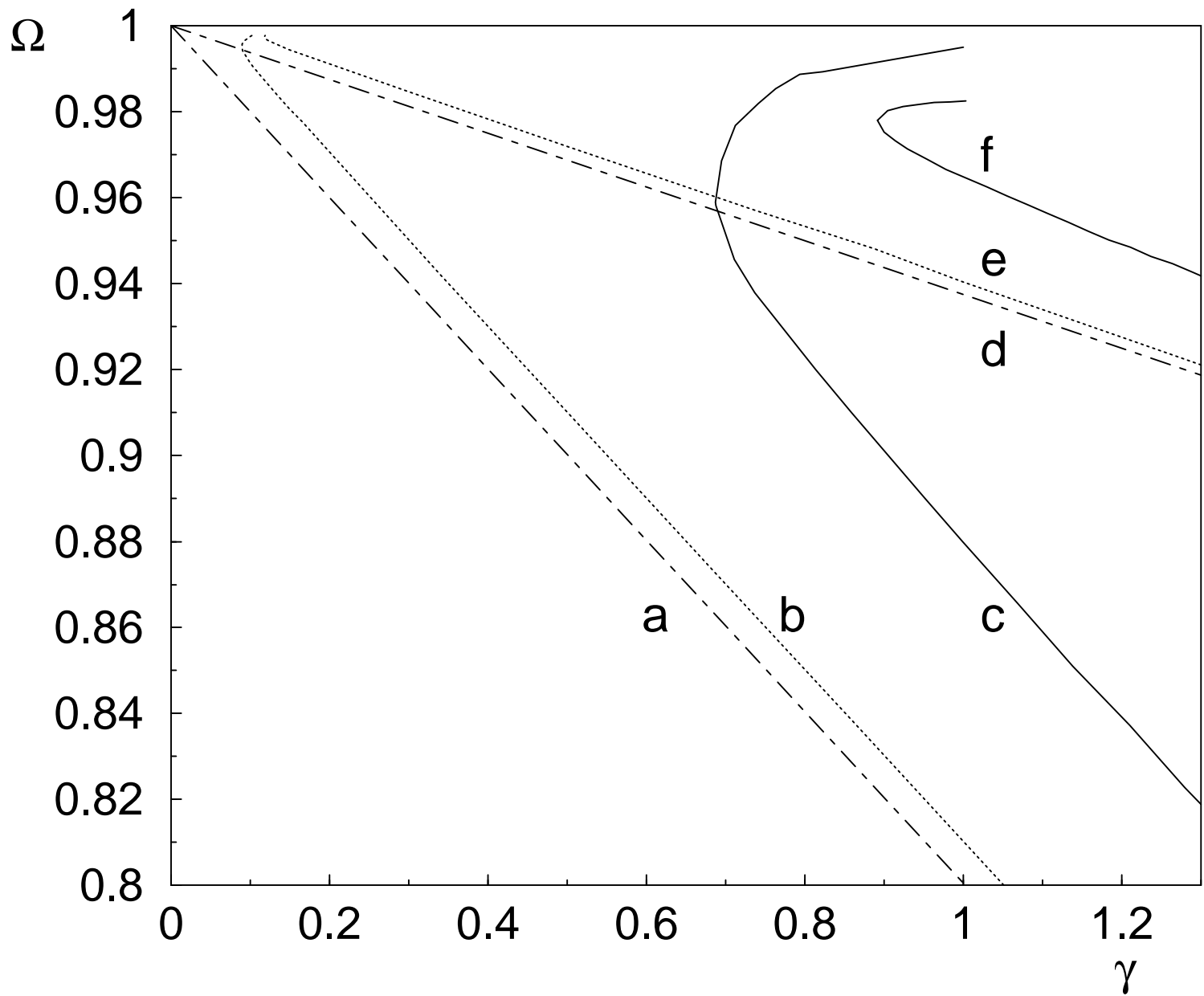
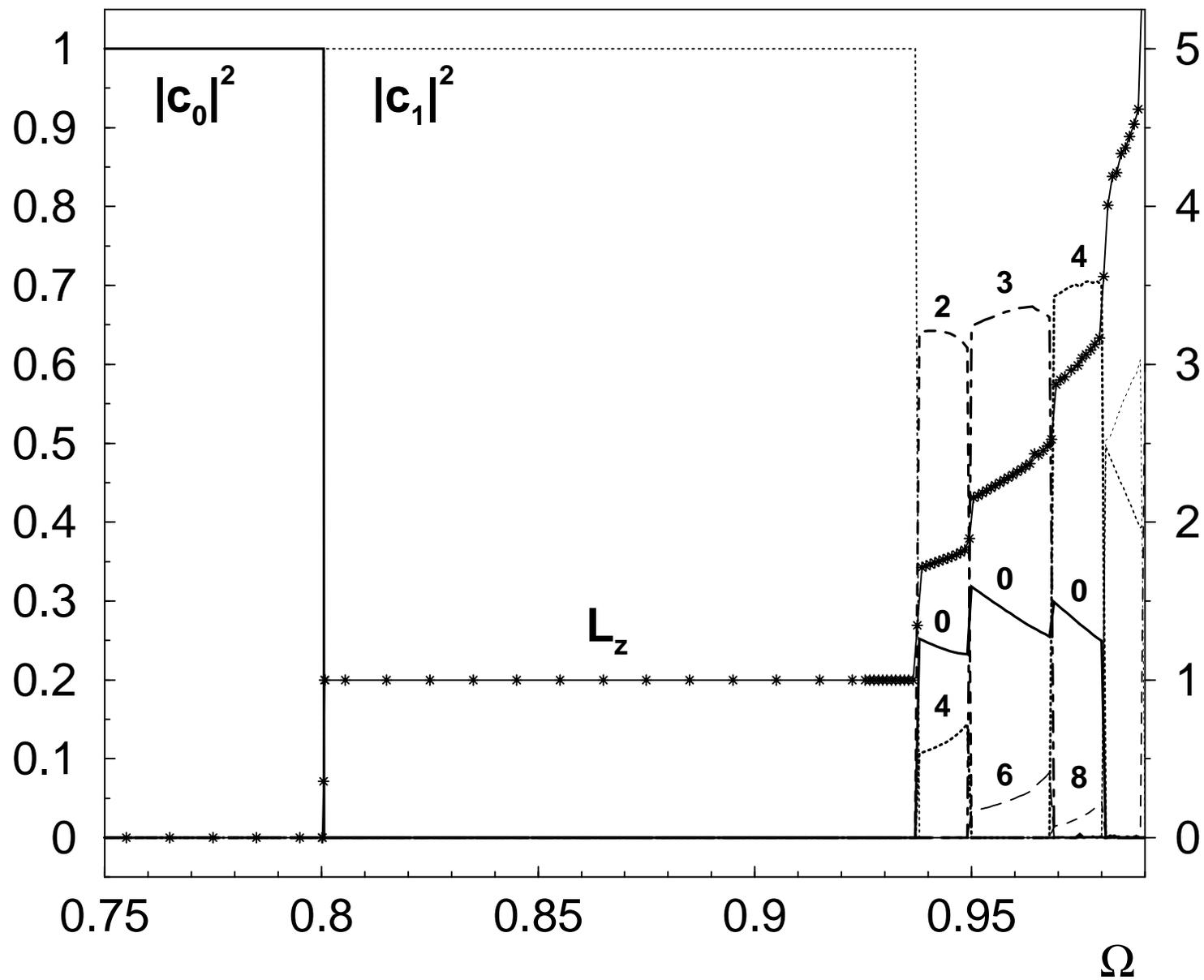
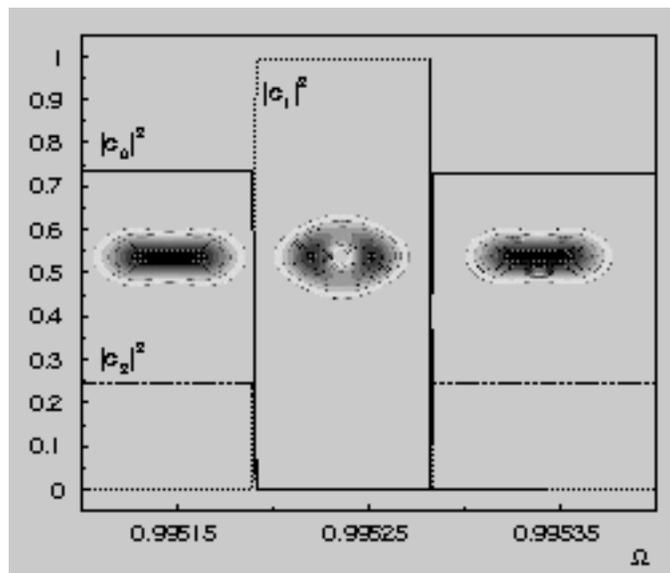
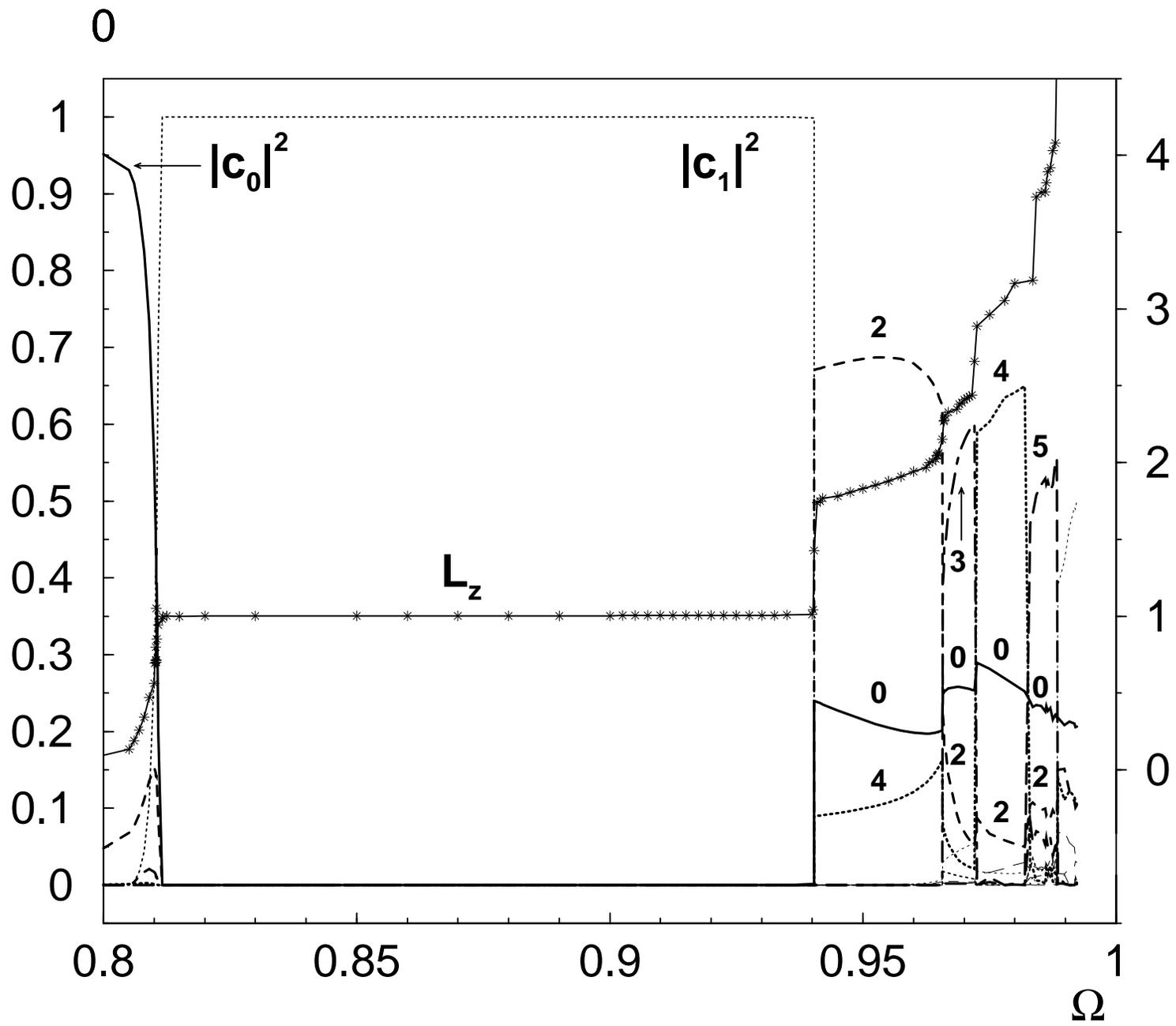
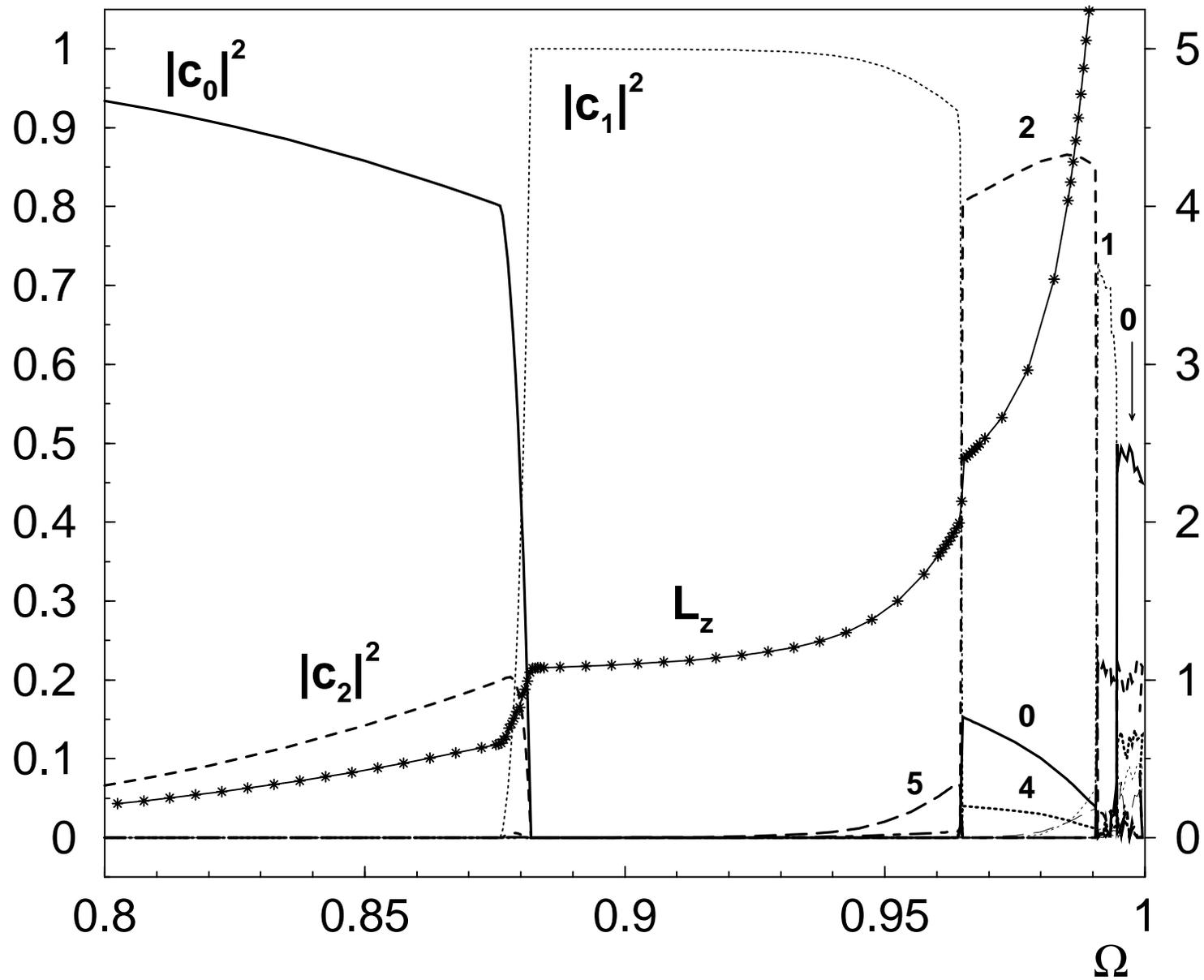
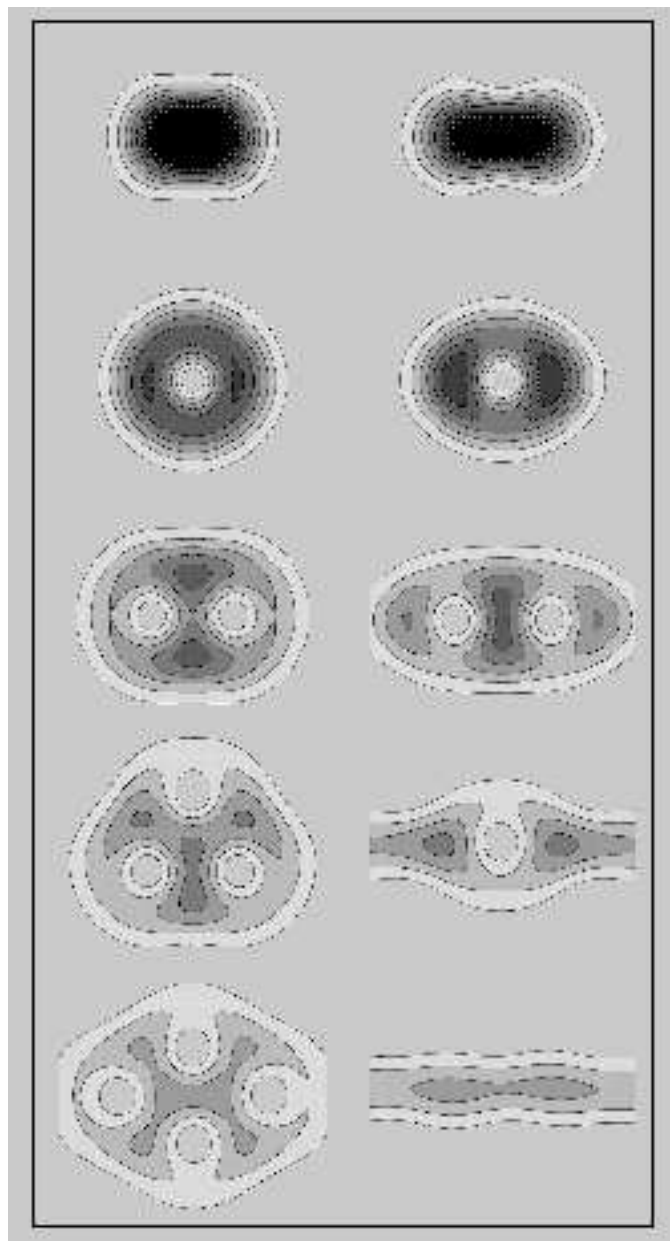
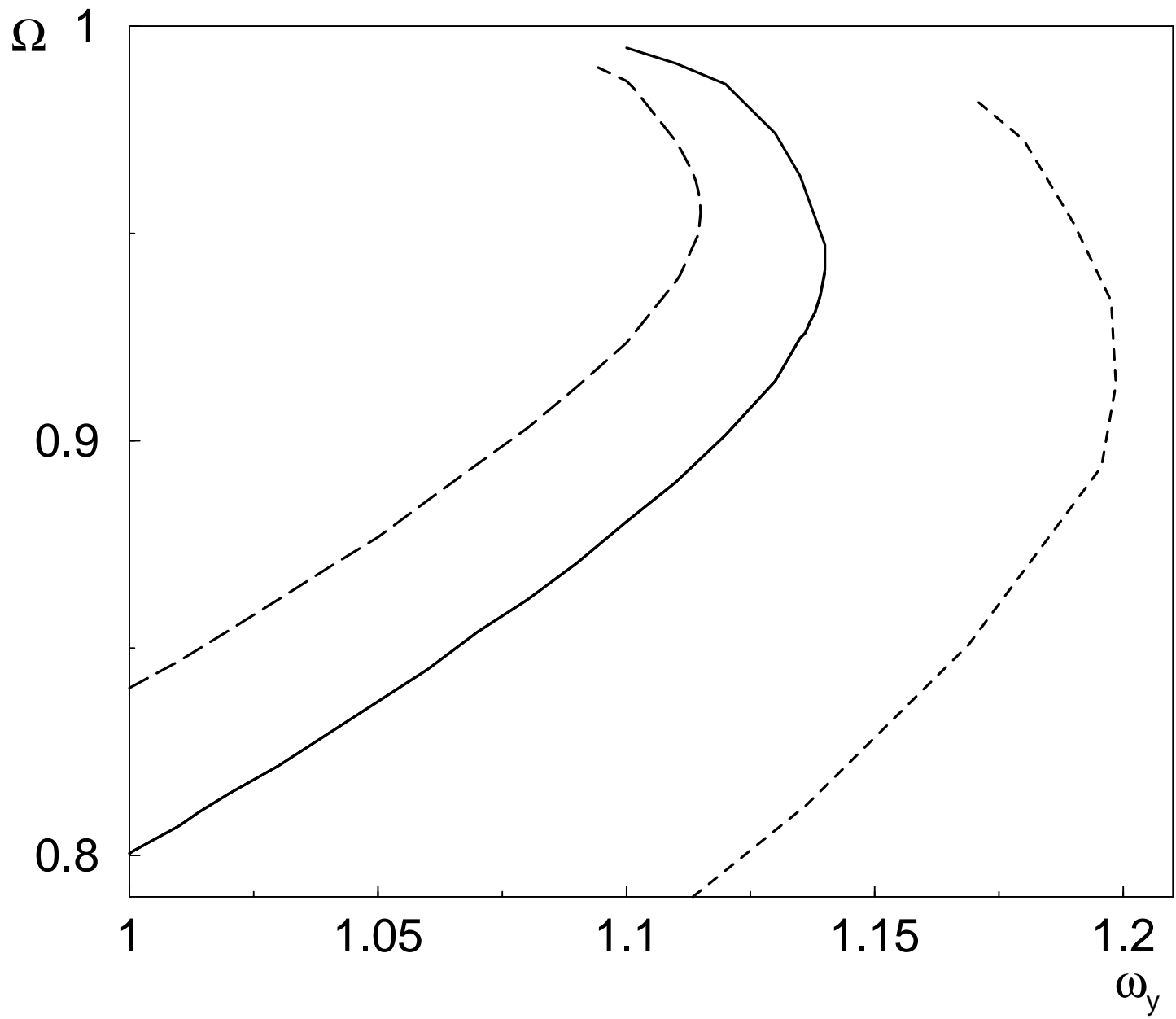
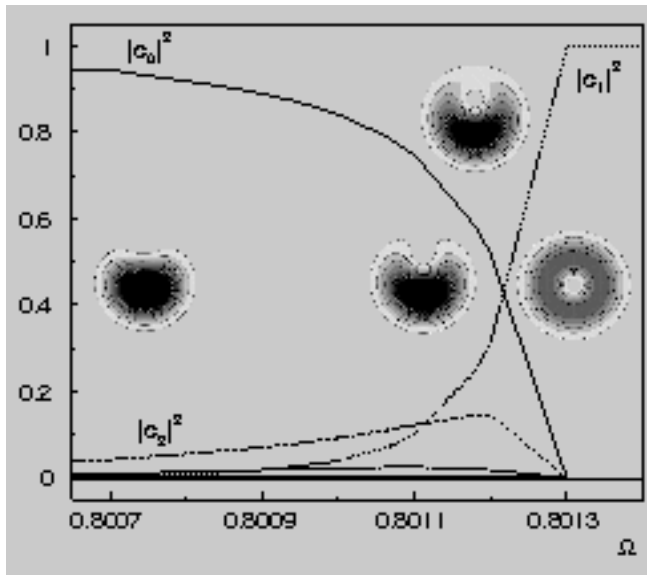
\end{document}